\documentclass[aps,pre,twocolumn,twocolumn]{revtex4}
\usepackage{color}
\usepackage{graphics,graphicx,epsfig}
\usepackage{epsf,epstopdf,wrapfig}
\usepackage{amssymb,amsfonts,amsmath}
\include{epsf}

\newcommand{\beqn}{\begin{eqnarray}}
\newcommand{\eeqn}{\end{eqnarray}}
\newcommand{\beq}{\begin{equation}}
\newcommand{\eeq}{\end{equation}}

\begin{document}

\title{Tiling solutions for optimal biological sensing.}

\author{A. M. Walczak}
\email[]{awalczak@lpt.ens.fr}
\affiliation{CNRS and Laboratoire de Physique Th\'{e}orique de l'\'{E}cole Normale Sup\'{e}rieure, Paris, France. }

\date{\today}
\linespread{1}

\begin{abstract}
Biological systems, from cells to organisms, must respond to the ever changing environment in order to survive and function. This is not a simple task given the often random nature of the signals they receive, as well as the intrinsically stochastic, many body and often self-organized nature of the processes that control their sensing and response and limited resources. Despite a wide range of scales and functions that can be observed in the living world, some common principles that govern the behavior of biological systems emerge. Here I review two examples of very different biological problems: information transmission in gene regulatory networks and diversity of adaptive immune receptor repertoires that protect us from pathogens.  I discuss the trade-offs that physical laws impose on these systems and show that the optimal designs of both immune repertoires and gene regulatory networks display similar discrete tiling structures. These solutions rely on locally non-overlapping placements of the responding elements (genes and receptors) that, {overall, cover space nearly uniformly}.
\end{abstract}

\maketitle

%--------------------------------------------------------------------------------
\section{Introduction}

A fascinating aspect of biological systems is the emergence of large-scale reproducible function from the small-scale molecular interactions between cellular elements (proteins, genes). Living systems, both whole organisms and molecular units, amaze us by the precision of their performance. How is this precision achieved under the physical constraints that biology must obey? One way to approach this question is to note that many biological systems display emergent behavior: macroscopic, stereotyped phenomena that cannot be explained merely by composing the properties of the system's underlying elementary, and intrinsically noisy, units. As we know from physics, nontrivial emergent behavior often results from interactions on different length, time, or energy scales. These effects are also ubiquitous in biological systems, for example in single cells expressing certain subsets of genes, in highly orchestrated multi-cellular programs such as development or the reliable response of {the} adaptive immune system against attacking pathogens. Examples of correlated phenomena have been extensively studied in statistical physics for the past century, leading to increasing our understanding of many-body interactions in condensed matter systems, as well as technological advances. 

Concurrently, recent advances in experimental technologies give us great insight into the functioning of biological systems both at the molecular, inner-cellular level, as well as the level of large scale functional systems in the organism and the behavior of large scale groups of animals. These technical developments allow us to make quantitative measurements of their constitutive elements and link it to their function. Trying to understand the functioning of these various systems, we see {the emergence} of common principles {governing} their behavior, despite the large biological differences, {their} functioning at different scales, and {their} fulfilling very different functions. In recent years physicists have become more interested in how physical principles are realized in cells. In the last decade, such an approach of taking inspiration from different biological systems (such as vertebrate development, chemotaxis, fly development, olfaction, visual processing) has proven very fruitful in proposing potential design principles (e.~g. error correction \cite{Hopfield1974, Ninio1975}, noise minimization \cite{Tostevin2007, saunders2009}, information transmission \cite{Tkacik2008a,Mehta2009,Walczak2010, Dubuis2013, Tostevin2009, Tostevin2010, deRonde2010}, acquiring information \cite{Vergassola2007}, speed and accuracy of decision making \cite{Siggia2013}, minimax strategies \cite{Celani2010}, evolvability \cite{Francois2004, Francois2007, Francois2010, Gerland2009}, optimization of resources \cite{Savageau1977, Scott2010, Klumpp2008}) that govern how physical laws are realized in living organisms. The lessons learned from these theoretical ideas have pushed the limits of experiments in concrete systems and often questioned our understanding of basic physical and biological processes.

Biological systems perform a function, limited both by the physical laws they must obey, as well as limited resources in the environment they find themselves in. Functioning efficiently and reliably in a given environment requires the matching of the statistical properties of the system to those of the environment, as has been discussed in the context of neuroscience \cite{Barlow1961, Laughlin1981}. If infinite sensing elements were available, the environment could be sensed up to the limits imposed by intrinsic physical noise. Of course this is not the reality of any biological system, where sensing and response must be fast and reliable and  natural trade-offs appear in the design of these systems. If we assume that the structure of biological systems makes it possible for them to reliably interact with their environment, we can attempt to understand which elements of form are linked to certain functions.

Here I will discuss two very different systems that perform two very different functions: genes and their regulatory proteins, inspired by regulation in developmental systems and the ensemble ({called} repertoire) of receptors expressed on the surface of immune cells. Generally, the goal when sensing is to cover the whole input space in such a way that each part of this space is well covered, given the constraints of limited resources. The detailed description and formulation of the goals of these two systems is very different, but similar trade-offs appear in these two different contexts. As we shall see, the optimal solutions {that} these two systems find are very similar, {although they are solutions} to very different problems that involve an adequate, yet different in their nature, response to their environments. In short, they both involve tiling the input space, be it the input concentration of a developmental gradient or the current distribution of antigens (elements of pathogens), with their sensory elements. I will concentrate on these two examples coming from my own work. However the idea of tiling by sensory systems has been wildly explored in neuroscience (where it is termed "lateral inhibition") and comes about naturally in information theory. I will mention briefly these two cases in the discussion. I will start by explaining the problem of interest in each of the systems and show how tiling solutions emerge in both cases before discussing the differences and similarities between {them}.

The work presented here is a review of work I did with different collaborators, all of whom have been exploring how sensory systems function. The gene regulatory system inspired by fly development was done in collaboration with Gasper Tkacik and William Bialek \cite{Tkacik2009, Walczak2010, Tkacik2011, Tkacik2012}.  I considered the question  of optimal immune repertoires with Andreas Mayer, Vijay Balasubramanian and Thierry Mora \cite{Mayer2015}.  In this review I chose to present only one aspect of the results obtained for these two systems - one that is common to both - tiling. The analysis in the original papers has many different perspectives that I do not discuss here.

\section{Gene regulation}

An important question in developmental biology is how the 
symmetry  at the early stages of the embryo is stably broken to create
structured organisms given the inherent cellular stochasticity. Examples of symmetry breaking are observed in development where the mother lays the foundation for gradients that are later translated to cell fates through a noisy signaling network. During development cells differentiate and start expressing different sets of proteins. The fruit fly (Drosophila melanogaster) is a model organism to study early embryonic development and cell differentiation \cite{Lawrence1992, Crauk2005}. The fly mother produces bicoid mRNAs, which are laid in the anterior of the egg. As the proteins translated from these mRNAs diffuse away from the pole they establish a decaying anterior-posterior protein gradient. Together with other maternal proteins (e.g. Nanos), Bicoid proteins locally regulate the expression a set of downstream "gap genes" (hunchback, kruppel, knirps and giant) to determine, among other features, the anterior-posterior (head to abdomen) axis. The gap genes control the expression of downstream genes (called pair-rule genes) that form very well defined stripes, which later lead to the formation of segments in the fly's body. The precise positioning and width of these stripes is essential for correct development. One of the puzzles of biology is how the position of the stripes can be controlled so accurately. All the positional information the fly embryo has is contained in the profiles of the maternal proteins. This information must be transmitted accurately in the different steps of gene expression, or the developmental plan will fail.

The task of transmitting the information about the concentrations of the maternal gradients is made more difficult by the fact that gene expression is a noisy process. On the molecular level, the interactions between genes and proteins occur by means of chemical reactions, which are probabilistic in nature. Furthermore the scarcity of the reaction products increases the intrinsic noise of the cell, requiring a stochastic framework \cite{Delbruck1940, Elowitz2002, Ozbudak2002, Raser2004, Kaern2005, Walczak2005a, Hornos2005, Walczak2010, Walczak2012}. There is typically one active copy of DNA per cell, a few copies of mRNA and tens to hundreds copies of a protein of a given species. The stochastic nature of gene expression has been confirmed experimentally \cite{Elowitz2002, Raser2004, Golding2005, Cai2006, Elf2007}.

Many elements of the gene regulatory network that regulate the expression of stripes have been mapped out. Owing to the experimental and theoretical advances of the last decade, we now have a good understanding of the molecular details of the basic forms of gene regulation. At the same time, our understanding of the basic components of molecular noise has increased. We can use this knowledge to go beyond the simple characterization of gene regulatory networks and ask whether we can identify the physical principles that govern the observed behavior of circuits. Specifically, can we understand why the early steps of cell differentiation in the fly embryo follow this specific pattern? How do the specific regulatory elements come together in space and time, and which parts of the regulatory process control which observed features? 

\begin{figure}
\includegraphics[width =  \linewidth]{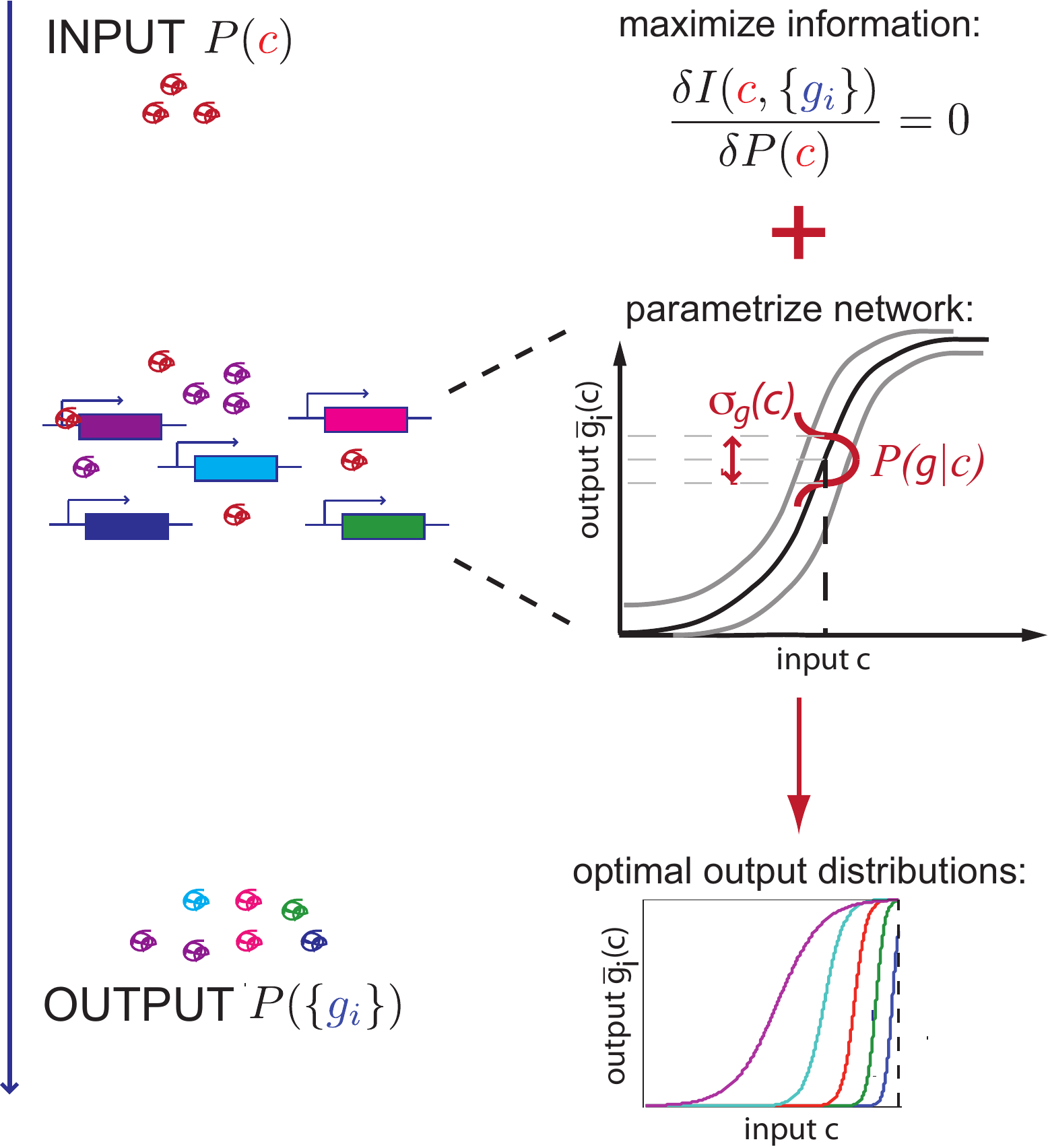}
\caption{Gene regulatory networks respond to input signals by producing output proteins.  Regulatory functions that optimize the information between the input and output require matching the statistics of the input distribution $P(c)$ with the properties of the network $P(\{g\}|c)$.  To do that we need to specify the nature of the regulation, which we assume is well described by the its mean regulatory function $\{g\}$ and Gaussian noise $P(\{g\}|c)\sim \mathcal{N}(\{g\}, \Sigma_c)$. The molecular biophysical properties of network are summarized in the form of the input/output function and the noise. For clarity of illustration this is portrayed on the example of one gene, but the picture generalizes to $L$ genes.  Optimizing information with respect to the input distribution and the properties of the network results in the optimal regulatory functions. The optimal functions were obtained assuming Hill regulatory functions, and are shown in Fig.~\ref{f-5genes}.}
\label{Fig1}
\end{figure}

In this initial stage of development, the continuous Bicoid concentration gradient gets translated into localized expression patterns of the gap genes: hunchback is only expressed in the first (anterior) part of the embryo, kruppel in the middle, giant and knirps in two sets of distinctly positioned stripes along the length of the embryo \cite{Jaeger2011}. Inspired by the regulation of the gap genes in early fly developed, Tkacik, Bialek and myself were interested in understanding the circumstances under which the expression of the target genes becomes localized \cite{Tkacik2009, Walczak2010, Tkacik2012}. Even these early stages of fly development include a number of complicated interactions between the target genes, so we started  by studying a simplified system where one continuous input (inspired by Bicoid) regulates $L$ independent downstream genes, inspired by the $L=4$ gap genes (see Fig.~\ref{Fig1}). We asked what are the regulatory interactions between the one input and $L$ output genes that maximizes the transmitted information between the input and outputs.

Information between the input and output is an intuitive concept that is also formally defined as mutual information in terms of the difference of entropies between the output distribution, $S[P(\{g\})$, and the conditional distribution of the output given the input, $S[P(\{g\}|c)]$ \footnote{Mutual information  is a symmetric quantity so it also can be defined as the difference of entropies of the input and the input given the output.} \cite{Shannon1948, Cover1991}:
\beq
I(c, \{ g \})=  \int dc P(c) \left( S[P(\{g\})] - S[P(\{g\}|c)] \right).
\label{info}
\eeq
 Entropy measures the uncertainty of a given distribution, $S[P({\rm x} )]=-\int d{\rm x} P({\rm x} ) \log P({\rm x} )$, where ${\rm x} $ is equal to $\{g\}$ and $\{g\}$ conditioned on $c$ in Eq.~\ref{info}. Knowing the entropy of the output distribution gives us a measure of our uncertainty of the output. However knowing the value of the output given a particular input further constraints our uncertainty. And precisely that reduction of our uncertainty about the output is the information we have gained by measuring the input. The subject of information in gene regulation has previously been reviewed and the interested reader can {refer to} \cite{Tkacik2011} for more details.
 
To make progress, we also need to characterize the regulatory network. We can assume that the conditional distribution of the output given the input that describes the regulation function is well approximated as a Gaussian with covariance $\Sigma(c)$ around the deterministic regulatory function. In general for $L$ interacting output genes $\Sigma(c)$ is the inverse covariance matrix of size $L\times L$ of the fluctuations in the expression levels {$\{g\}$} at fixed input {$c$}. However in the specific case discussed in most of this review of $L$ non-interacting genes, $\Sigma(c)$ only has diagonal entries coming from the variance in its own output and the global input $c$, $\Sigma_{\rm ij}(c)=  \sigma^2_{i}(c) \delta_{i,j}$. The Gaussian assumption effectively reduces the problem of describing the input and output probability distributions, $P(\{g\})= \int dc P(\{g\}|c) P(c)$ to knowing the input distribution and the parameters of the $L$ dimensional Gaussian distribution{, i.e. its mean $\{\bar g(c)\}$ and covariance matrix $\Sigma(c)$}. Since this is rather repetitive to draw for $L$ noninteracting genes, Fig.~\ref{Fig1} depicts the parametrization of the network for one output gene. 

We can now use our knowledge of the regulatory interactions and noise properties in gene regulation and account for all the biophysical constraints by parametrizing the variance and the mean regulatory functions. We chose a basic thermodynamic Hill model for regulation (see below) and assumed that most of the noise comes from the random production of proteins (Poisson noise - first term in Eq.~\ref{sigmaexplicitly}) and the diffusion limited switching of the genes (a Berg-Purcell term \cite{Berg1977} - second term in Eq.~\ref{sigmaexplicitly}):
\begin{equation}
\sigma^2_{i}(c)= \frac{1}{N_{\rm max}} \left[ \bar g_{\rm i}(c)+ c c_0 \left({{d\bar g_{\rm i}(c)}\over{dc}}\right)^2 \right],
\label{sigmaexplicitly}
\end{equation}
where $N_{\rm max}$ is the maximum number of independent molecules that are made from gene $i$ and $c_0=N_{max}/{Da\tau}$ is the characteristic concentration scale composed of the characteristic length and timescales for regulation (the diffusion constant $D$, the size of the target binding site $a$ and the input integration time $\tau$).  By doing this we effectively parametrize the variance by the mean input-output relation,  $\bar g_{\rm i}(c)$.  The Hill functions that describe  $\bar g_{\rm i}(c)$ are sigmoidal smooth monotonic functions of the input concentration
\begin{equation}
\bar g_{\rm i}(c) = {{c^{h_{\rm i}}}\over{K_{\rm i}^{h_{\rm i}} +  c^{h_{\rm i}}}} 
\end{equation}
parametrized by the concentration which results in half-maximal expression of the gene, $K_{\rm i}$, called the dissociation constant, and the steepness of the regulatory function $h_{\rm i}$, which is linked to the cooperativity of the molecular reactions involved in regulation. The form of this expression can be derived from thermodynamic arguments with $K_{\rm i}=\exp(-F/k_BT)$ where $F$ is the the free energy of binding per input molecule. The sign of the cooperatively parameter $h_{\rm i}$ differentiates between activation ($h_{\rm i}>0$) and repression ($h_{\rm i}<0$) of the target gene by the input, and its value extrapolates between linear regulation (($h_{\rm i}=1$) and threshold switching ($h_{\rm i}\rightarrow \infty$). Each gene responds by producing a differentiable output in a limited input concentration range $\sim K_{\rm i}/h_{\rm i}$ set by the cooperativity coefficient and measured in units of $K_{\rm i}$ around the concentration midpoint given by the dissociation constant $K_{\rm i}$. Below that range the gene is essentially off, and above it has saturated its expression. More details of the parametrization can be found in \cite{Tkacik2009, Walczak2010, Tkacik2011}. In summary, this parametrization allows us to describe the properties of the network in terms of two parameters for each gene: the dissociation constant, $K_{\rm i}$, which describes the positioning of the gene in the input concentration range and the cooperatively function which sets the range of inputs the gene is responsive to. To find the optimal network we must find the optimal values of these parameters.

Biological systems must obey a number of constraints, including paying a cost for producing molecules. We can now optimize mutual information with respect to the input distribution given this molecular constraint that we include by limiting the range of input molecules (technically imposing an upper integration bound for input concentrations) and demanding the distribution is normalized:
\begin{eqnarray}
\frac{1}{\delta P(c)} \Big[ S[P(\{g\})]-\int dc\; P(c) S[P(\{g\}|c)] +&&\\
\lambda \int dc\;P(c)\Big]=0.&&
\end{eqnarray}
The expression is easily minimized, because we have parametrized $P(\{g\}|c)$ as a Gaussian defined explicitly in terms of biophysical rates of the problem. The optimal input distribution {$P^*(c)$} is expressed in terms of the {uncertainty of the input $c$ given the outputs $\{g\}$, given by the variance of the posterior $\sigma_c^2 (\{g\})$}: 
\begin{eqnarray}
{1\over {\sigma_c^2 (\{g_{\rm i}\})}}& =& \sum_{{\rm i,j}=1}^L 
\left[
{{d\bar g_{\rm i}(c)}\over{dc}} {[\Sigma(c)^{-1}]}_{\rm ij} {{d\bar g_{\rm i}(c)}\over{dc}}
\right]
{\Bigg |}_{c = c^*(\{g_{\rm i}\})} \\
&=&\sum_{{\rm i}=1}^L 
\left[
\left({{d\bar g_{\rm i}(c)}\over{dc}}\right)^2 \frac{1}{\sigma^2_{{\rm i}}(c)}
\right]{\Bigg |}_{c = c^*(\{g_{\rm i}\})},
\label{sigmaeqn}
\end{eqnarray}
{through}
\begin{eqnarray}
P^*(c) &\propto& {1\over{\sigma_c(\{\bar g_{\rm i} (c)\})}}\\
&=& {1\over {Z}}
\left[{1\over{2\pi e}} \sum_{{\rm i,j}=1}^L
{{d\bar g_{\rm i}(c)}\over{dc}} {[\Sigma(c)^{-1}]}_{\rm ij} (c) {{d\bar g_{\rm j}(c)}\over{dc}}
\right]^{1/2},
\label{Popt}
\end{eqnarray}
with the normalization given by
\begin{equation}
Z = \int_0^{c_{\rm max}} dc\, \left[{1\over{2\pi e}} \sum_{{\rm i,j}=1}^L
{{d\bar g_{\rm i}(c)}\over{dc}} {[\Sigma(c)^{-1}]}_{\rm ij} (c) {{d\bar g_{\rm j}(c)}\over{dc}}
\right]^{1/2}   .
\label{Z1_multiple}
\end{equation}
The last line in Eq.~\ref{sigmaeqn} explicitly gives the expression for $L$ non-interacting genes, whereas the other expression keep the general form for possibly interacting genes. Given the optimal input distributions we can calculate the the optimal information of the system
\begin{equation}
I^* (c;\{g_{\rm i}\}) = \log Z .
\end{equation}
  
\begin{figure}
\includegraphics[width =  \linewidth]{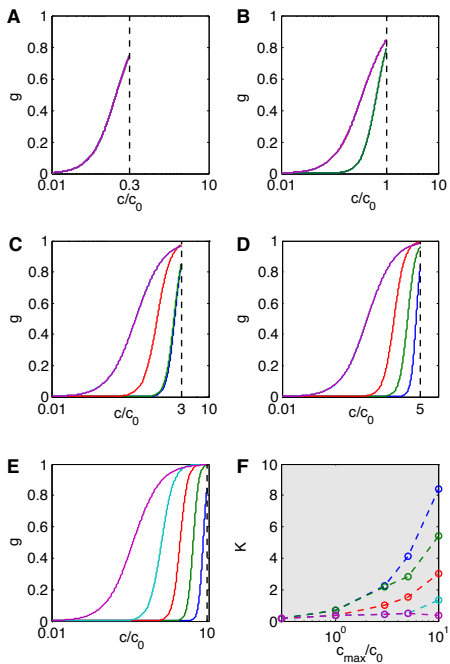}
\caption{The most informative regulatory functions $\{g_1(c),\dots,g_5(c)\}$ (shown in different colors) for a gene regulatory network with one input $c$ regulating $L=5$ non-interacting output genes, for increasing ranges of the maximum input concentrations $c_{\rm max}$ (A--E).  At very small values of $c_{\rm max}$ all $L$ genes have the same regulatory function making the readout completely redundant. For increasing values of $c_{\rm max}$ the redundancy of the readout is lifted, as successive genes individually cover particular subranges of the input concentration. (F) The emergence of the tiling solution in terms of the gene {dissociation} constants $K_{rm 1}, ..., K_{rm 5}$ as a function of $c_{\rm max}$. We assumed Hill regulatory functions and Berg-Purcell and small molecule noise dominate protein fluctuations, as described in \cite{Walczak2010}.}
\label{f-5genes}
\end{figure}

The details of this calculation can be found in \cite{Walczak2010}. The optimization problem requires finding the optimal parameters of the regulatory functions ($\{K_{\rm i }, h_{\rm i}\}_{\rm i =1,..,L}$) that maximize $\log Z$, whose form is determined by the noise. The optimal solutions are a result of the balance between the two sources of noise in Eq.~\ref{sigmaexplicitly}: the input noise coming from fluctuations in regulatory protein concentrations (second term in Eq.~\ref{sigmaexplicitly}), and the output noise caused by the small number of produced proteins $g_{\rm i}$ (first term in Eq.~\ref{sigmaexplicitly}). At small input concentrations the fluctuations from an unreliable readout of $c$ dominate and push the solutions to have higher values of $K_{\rm i}$, whereas the need to distinguish different levels of outputs reliably decrease the steepness of the regulatory functions and forces them to use also the smaller concentration ranges, decreasing $K_{\rm i}$ and $h_{\rm i}$. The actual parameters of the regulation function need to be optimized numerically and for large concentration ranges of input molecules we obtained the characteristic optimal regulatory functions as shown in panels B--E of Fig.~\ref{f-5genes}. The different panels show the solutions for increasing input concentration ranges. In these solutions each gene is effectively regulated in a finite localized input concentration regime. For small concentrations the first gene is expressed and when it saturates, the second gene is expressed. This trend continues and the gene expression domains {\it tile} space. 

The tiling solution is the most informative solution when the concentration range of input molecules is large. When the concentration range of the input molecules is small (see Fig.~\ref{f-5genes}~A), the optimal solution consists of all genes making the same readout of the input by having exactly the same regulatory function. In this case the genes no longer tile space, but repeat the same measurement to minimize the error coming from reading out small concentrations. The transition from one regime to another is continuous as a function of input range concentration and information.
{This can be explained in terms of the dissociation constants of each gene, which give the concentration value at which a gene is expressed at half maximum.}
Either there is enough concentration range to use the discrete gene readout ---{in} this case the genes have different values of the  {dissociation} constants; or it is better to attempt a reliable readout in one concentration regime, and  {the dissociation} constants collapse. The transition from the non-tiling to tiling regimes in terms of the  {dissociation} constants is shown in Fig.~\ref{f-5genes}~F. Figure~\ref{f-5genes} also shows that the tiling of genes is gradual and the redundancy is lifted {one gene at a time} as the input concentration range increases. 

\section{Immune receptors}

Let me now leave gene regulation and turn to a completely different system - the adaptive immune repertoire. After this short presentation, I will return to the our information optimal gene regulatory network and compare the characteristics of these two very different biological systems.

The role of the immune system is to protect the organism from the many pathogenic threats it constantly encounters. To fulfill this role, it must be prepared to identify a great variety of unknown challenges, including ones it has never been exposed to. It must thus maintain a diversity of specialized cells, each specific to particular challenges, but which together cover the full array of potential threats. These cells are called B and T-cells and the particular receptors responsible for recognition on each of these specialized cells are generated in an essentially random manner \cite{Janeway}. Yet together these receptors form a diverse repertoire that allows the immune system to fulfill its function of recognizing pathogens exceptionally well. Since not all threats are equally likely, the immune repertoire adapts to the changing pathogenic environment, at the same time keeping a memory of past infections. The diversity of the composition of the immune repertoire emerges as a self- organized process, stimulated by interactions with the environment.

Receptor proteins on the surfaces of these cells interact with pathogens, recognize them through specific binding and initiate the immune response.  The interaction between pathogen proteins and receptors is based on the binding of two polypeptides (one being part of the receptor, the other being part of the pathogenic protein) and is specific, yet degenerate: a single receptor is able to recognize more than one pathogen peptide (antigen) and, conversely one antigen can bind to more than one receptor. How the highly dimensional space of pathogenic peptides is covered by receptors is a particularly challenging example of a covering problem.

The immune response is controlled by many factors on many scales, however it is initiated when a B or T-cell receptor successful recognizes an element of a foreign pathogen, called an antigen. In our approach, presented in \cite{Mayer2015}, we decided to focus on this part of the puzzle and ask how should immune receptors be distributed in order to minimize the harm from infections given a fixed, static antigenic environment. Additionally, antigens and receptors have a limited number of encounters, since it takes time for a receptor and antigen to meet given the finite concentrations of both and the size of the organism. This imposes a constraint on the efficiency of recognition. By formulating the problem in this way, we simplified it to a static version of a covering problem with limited resources. However even in this simple formulation the exact meaning of most of the used terms needs to made precise. 

\begin{figure}
\includegraphics[width =  \linewidth]{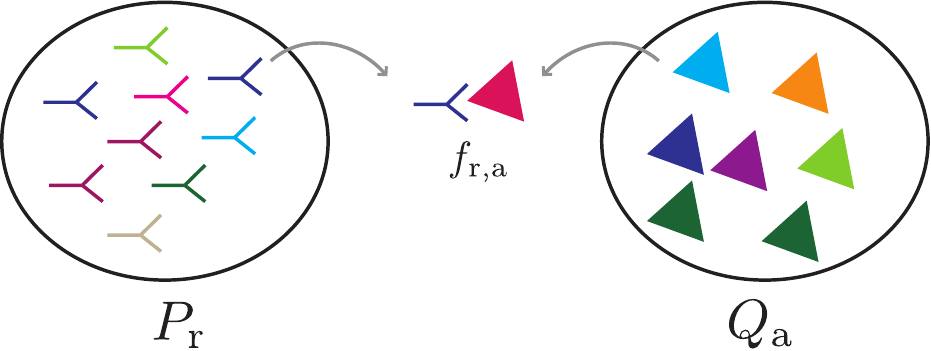}
\caption{The simplified recognition problem in the adaptive immune system: receptors from the repertoire distribution $P_{\rm r}$ recognize antigens from the environment distribution $Q_{\rm a}$ with a  cross-reactive recognition probability $f_{\rm r, a}$.}
\label{immcartoon}
\end{figure}

Since recognition is triggered by binding of receptors and antigens, we can consider the problem in an effective recognition space. Both types of molecules (antigens and receptors) live in this space and they recognize each other if the distance between them in this space is small. The idea of recognition space is similar to shape space \cite{oster-1979} which has been very useful for decades for describing effective antigen-receptor interactions. We do not need to further parametrize the space to describe this interaction (for example as has been done using string models, where both molecules are taken to be strings of effective amino acids with effective physical and biochemical properties and their similarity is measured in terms of a Hamming distance between the strings), however this specific picture is a helpful concrete example of the type of effective recognition space we have in mind. Since we are thinking about a static picture, the receptor repertoire is described by a probability distribution $P_{\rm r}$ and the fixed ensemble of antigens  by $Q_{\rm a}$, as depicted in Fig.~\ref{immcartoon}. A receptor and antigen can meet and recognize each other with a cross-reactive recognition probability $f_{\rm a, p}$. This function accounts for the fact that one receptor can recognize many pathogens and, conversely, one antigen can be recognized by many receptors. Given the recognition probability, the probability of an immune response from an encounter {of a random receptor} with a given antigen, $a$, is $\tilde P_{\rm a} = \sum_{\rm r} f_{{\rm r},{\rm a}} P_{\rm r}$. 

Since we are interested only in the consequences (recognition or not-recognition) of encounter events, we choose to measure time in the mean number of encounters $m$. 
The recognition events are random and Poisson distributed in  {$m$}. The limitation on the number of encounters can also be understood in terms of finite sampling of the {receptors by the antigen}. As the number of encounters increases {while} the antigen remains unrecognized, the effective cost of the infection increases {according to a function} $F_{\rm a} (m)$, due to the damage caused by the potentially proliferating antigen to the tissues of the organism. Therefore to obtain the total harm caused by a given antigen we need to integrate the effective cost over the number of encounters weighted by the distribution of successful recognition encounters:
\beq
\bar F_{\rm a}( P_{\rm r}) = \int_0^{+\infty} dm \, F_{\rm a}(m) \, \tilde{P}_{\rm a}  e^{-m \tilde{P}_{\rm a}}.
\eeq
Finally, the overall cost to the organism needs to take into account the costs from all the antigens:
\beq
\mathrm{Cost}(\{ P_{\rm r}\})=\sum_{\rm a} Q_{\rm a} \bar F_{\rm a}( P_{\rm r}).
\eeq
This cost accounts for the trade-off between a having to distribute many receptors given a finite number of encounters in such a way that the total harm caused by infections increases with time. Given a fixed antigen distribution, $Q_{\rm a}$ we find the optimal distribution $P_{\rm r}$ of receptors that minimizes this cost. 

\begin{figure}
\includegraphics[width =  \linewidth]{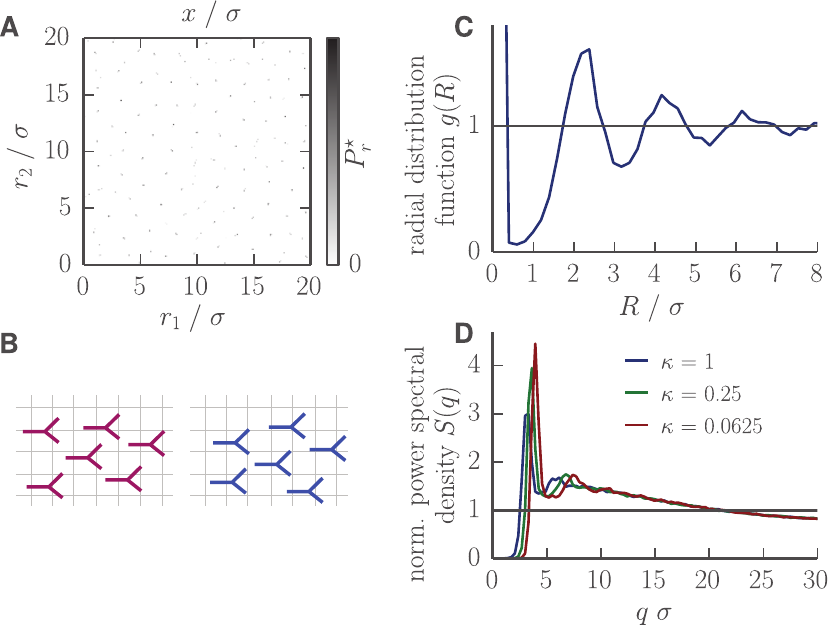}
\caption{Optimal receptor distribution $P^*_{\rm r}$ 
      for two-dimensional random environments (A). The antigenic landscape $Q_a$ is generated randomly from a log-normal
distribution with coefficient of variation $\kappa=1$. (B) A cartoon representation of the main characteristics of the optimal receptor distribution: locally the receptors are placed randomly and are non-overlapping, whereas at large distances they uniformly tile space in a way that two locally distinct patterns are indistinguishable on large scales. The lines guide the eye to compare the two patterns. These properties are well quantified by analyzing the tiling patterns using correlation functions (C-D).  (C) The radial distribution function of $P^*_r$, $g(R)$, has an exclusion zone at small distances around each peak, with a periodic pattern at large distances characteristic of a local tiling pattern.
        (D) The normalized power spectral density $S(q)$ of $P^*_r$ for different values of the coefficient of variation of the distribution function of the environmental, $\kappa$, quantifies the heterogeneity of the antigenic landscape at different scales. Fluctuations average out at large scales (small $q$) to uniformly cover space, whereas locally (large $q$) there are many possible placements for the receptors (non-zero $S_q$).}
\label{immunerep}
\end{figure}

The details of this optimization, as well as some analytical intuition gained from limiting cases is discussed in detail in \cite{Mayer2015}. The interesting results for the purpose of the current discussion are best seen in the numerical results recounted in Fig.~\ref{immunerep}~A for a two dimensional random antigen distribution. We see the optimal receptor distribution tiles space in a random way: the distribution is discrete with receptors positioned as individual non-overlapping peaks in recognition space. To quantify this pattern in more detail we can look at the radial distribution function as a function of distance between the receptor positions (Fig.~\ref{immunerep}~C). We see a strongly repelling core at small distances, that forbids the placing receptors too close to one another and characteristic regular peaks at larger distances that indicate likely positions of the receptors. Analyzing the structure function, confirms our intuition that the precise placement of the receptor is not important (Fig.~\ref{immunerep}~D). The structure factor at long wavelengths corresponding to short inter-receptor distances  {goes to 1}, as  it does in a {liquid or disordered glass}. However at large scales the pattern is completely reproducible, as quantified by  the structure factor {going} to zero at short wavelengths. Cross-reactivity allows one receptor to cover all the antigens within a given range and results in the hard core, whereas needing to protect against even the rare but potentially dangerous antigens requires a thorough coverage of the whole space. 

A biological implication of this type of receptor distribution is the fact that two individuals seeing roughly the same antigen environment can have dramatically different optimal repertoires aimed at targeting this specific set of antigens, while both manage to attain complete covering of the antigens, {as illustrated by Fig.~\ref{immunerep}~B}. Two individuals can see two slightly different versions of the same environment simply due to sampling different antigens. Further discussion of the immunological implications can be found in the original paper \cite{Mayer2015}.

The appearance of the tiling pattern in the optimal solution depends on the value of cross reactivity. This is most easily seen in the limiting case of taking both the antigenic environments and cross-reactivity to be Gaussian functions \cite{Mayer2015} - the localized peaked receptor distribution is optimal only when the variance of the cross-reactivity $\sigma$ is more than $\sqrt{2}$ times larger than the variance of the pathogen distribution $\sigma>\sqrt{2} \sigma_Q$. The transition is continuous: the variance of the receptor distribution decreases until it becomes a delta function.

\section{Connection to real systems}

The reasoning presented above finds the form of idealized optimal regulatory networks and repertoires. The role of such an approach is not necessarily to explain the detailed form of a given biological design. In fact in both cases, as is especially clear in the case of the immune repertoire, we have greatly simplified the problem and not taken into account important signaling and spatial dependencies of the system. Our goal in both case was to learn some general properties and understand which ingredients lead to what type of characteristics. However despite this great simplification, we do reproduce certain broad features of both gene regulation and immune repertoires. In the first case, the separation of the input space into domains of specific genes resembles the set-up of gap genes in early fruit development. As we discussed in detail in \cite{Walczak2010} adding interactions between the output genes, such as exist between certain gap genes, allows us to reproduce similar domains of expression bounded from low and high concentration values by domains where the gene does not produce proteins. 

Direct comparison to experiments is harder in the case of immune repertoires since our predictions are in the effective recognition space that has not yet been mapped out experimentally, whereas experiments give us receptor sequences. However, as noted above, this approach results in certain concrete predictions. Specifically, we expect the repertoire of two individuals to be different even if they are exposed to the same environment. Controlling the environment for humans is impossible, but detailed studies of {shared sequences between individuals} show no more overlap than expected by chance \cite{Murugan2012, Elhanati2014, Venturi2006, Venturi2008}. Zebrafish B-cell repertoire studies also showed no correlation between the environment and repertoire \cite{Mora2010, Weinstein2009}.  Another corollary of our predictions is that receptors should form well separated clusters. Analysis of the amino acid sequences of CDR3 regions in zebrafish B-cell repertoires showed that the sequences cluster into a relatively small number of attractors that are mostly different from the genetic templates \cite{Mora2010}. This hints that a tiling solution in effective recognition space is not incompatible with the data.

\section{Discussion}

\begin{figure}
\includegraphics[width = 0.5 \linewidth]{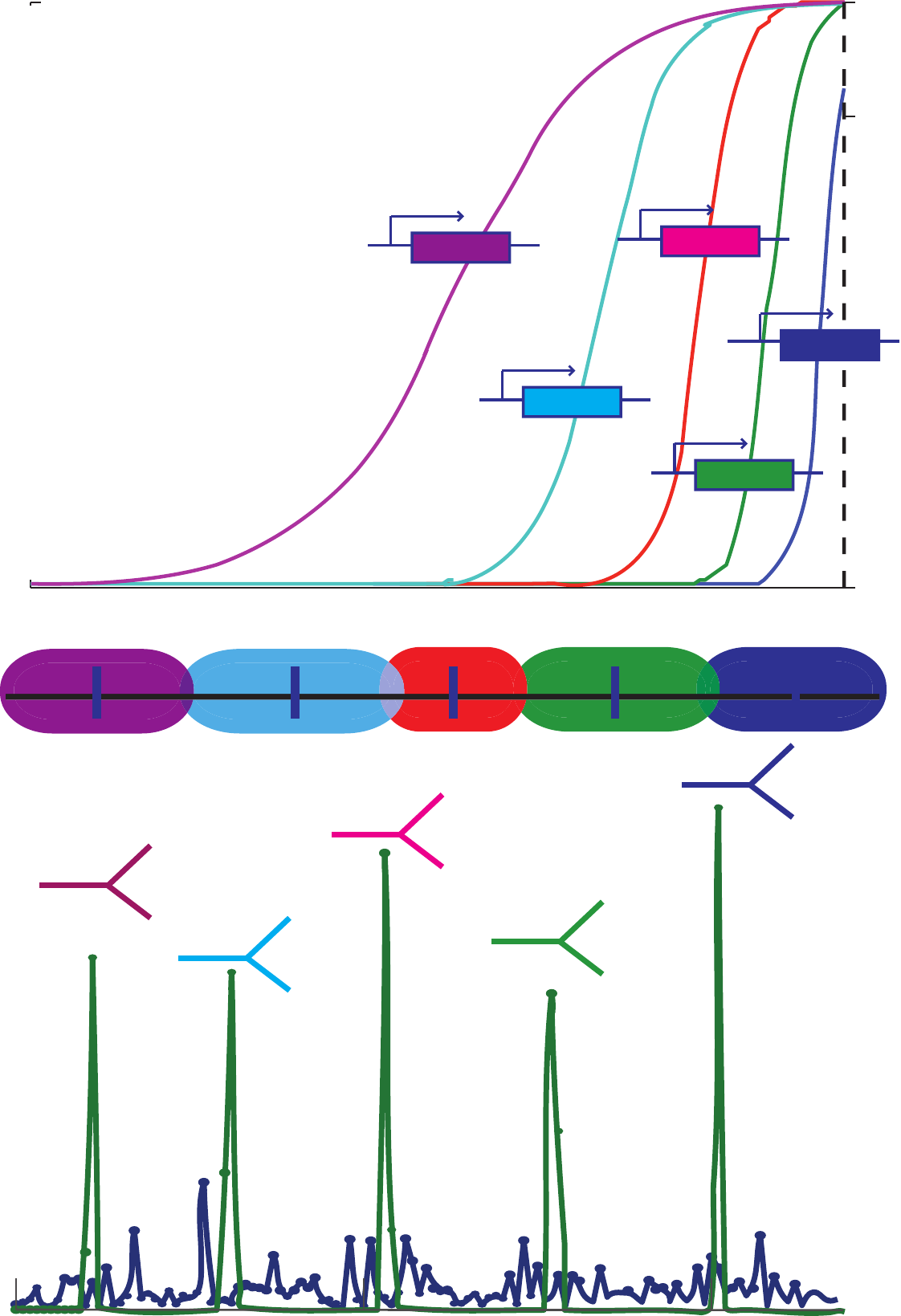}
\caption{The tiling of space by sensory elements emerges as the optimal design in two biologically different settings: gene regulatory networks and immune repertoires. The precise position of the tiling elements varies and is not uniform. It  is dictated by the matching between the properties of the environment and the response elements. Yet the whole space is fully covered (middle cartoon). The solution  that maximizes information transmission between the input and outputs of gene regulatory networks tiles the concentration range (top) and the cost to the organism is minimized when the repertoire tiles the recognition space between receptors and antigens (bottom). The details of the gene regulatory optimization are discussed in the caption to Fig.~\ref{f-5genes}.  The optimal receptor distribution $P^*_r$ (green line) in one dimension is shown for a random environment with an antigenic landscape $Q_a$ (blue line) generated randomly from a log-normal
distribution with coefficient of variation $\kappa=1$. The optimal repertoire is peaked, however the coverage of antigenic space $\tilde P^*_a=\sum_r f_{r,a}P^*_r$ is close to uniform.}
\label{summarycartoon}
\end{figure}

In the case of the two systems considered here, the initial formulation of the problem is very different. In both cases we looked for optimal solutions, however they are optimized for very different quantities. The gene regulatory network is optimized for information transmission between the input and output - a common assumption in many sensing systems \cite{Laughlin1981, Fairhall2001, Vergassola2007, Andrews2007}. {By contrast,} the immune repertoire is optimized to guarantee the least costly response - to allow the immune system to respond to a diversity of antigens with the smallest delay weighted by the potential harm of not responding. Nevertheless in both problems there is a trade-off given by the limitations of the system. In gene regulation the cell does not have infinite input molecules at its disposal, forcing it to distribute its response genes. In the immune system the number of encounters between antigens and receptors is {finite},  {limiting} the number of potential recognition events. These two trade-offs leads to similar types of solutions that show the same kind of characteristic features on short and large scales.

Naturally the biological nature of the gene regulatory and repertoire problem is very different. Additionally even in terms of their formal theoretical description the systems are very different. The gene regulatory problem includes placing discrete genes in a continuous space, making the problem discrete from the beginning. In the case of the repertoire the space is continuous and the discrete distribution emerges by itself.

Despite these differences the gene regulatory and immune repertoire show very similar tiling structures at different scales (see Fig.~\ref{summarycartoon} for a direct comparison). Locally we observe clustering in both cases. The optimal repertoire has the structure of discrete peaks although the probability of having the receptors at a given position in recognition space is a priori continuous. The receptors are discrete non-overlapping entities. Similarly the range of input concentrations in which a given gene is expressed, and in which we can discriminate the input concentration by measuring the {output}, is locally limited. Reducing {that} input concentration range results in the clustering of { the regulatory properties of the output} genes, and the overall range {of inputs they respond to}. In both systems we see this clustering of function on small scales: some areas of recognition space or input concentration are directly covered by a receptor or a given gene, whereas nearby areas are not. Yet on the large scale in both cases the whole space is nearly uniformly covered by genes or receptors, as seen explicitly from the close to uniform coverage of the antigens by the receptor distribution, $\tilde P_{\rm a} = \sum_{\rm r} f_{{\rm r},{\rm a}} P_{\rm r}$ \cite{Mayer2015}. So at large scales we observe the characteristic tiling of all of space by the total distribution of receptors, which leaves no part of effective space uncovered.  

The analogy depicted in Fig.~\ref{summarycartoon} is made more explicit in terms of the parameters of the two problems. The specific placement of the receptor corresponds to the concentration at half maximum expression of the gene (${K_{\rm i}}$) - the placement of the gene in input concentration space. But just as cross-reactivity ensures that areas of recognition space where there are no receptors remain covered,  $\sim K_{\rm i}/h_{\rm i}$ sets the range of input for which a given gene differentially responds to input concentrations.

This structure of local clustering and global tiling of space depends on the ratio of cross-reactivity to the size of the space for immune receptors and the value of the maximum input concentration scaled in natural units of concentration for gene regulation. Natural units of concentration result from the characteristic physical timescales of regulation (diffusion constant $D$, signal protein integration time $\tau$, the maximum number of independent molecules of the output $N_{\rm max}$ and typical size of binding site $a$), $c_0=N_{max}/{Da\tau}$. If we consider a relatively large effective space (whether it is the input concentration range or recognition space) the optimal solution will consist of $L>1$ genes and  $L>1$ receptors that use cross-reactivity to cover the whole pathogenic space. If we now decrease this effective space, the optimal genetic solutions will  reduce to one gene and similarly the whole recognition space can easily be covered by one receptor. The discrete structure of the optimal immune repertoire distribution depends on the small scale noise. If there was no intrinsic variability at small scales the optimal distributions would be continuous since nothing would differentiate between particular points in space. This effect can be seen in the limiting case of Gaussian antigen distributions where  {there is no small-scale noise at all} and the optimal receptor distribution is also Gaussian (given a Gaussian cross-reactivity function). 

The fact that solutions that optimize information in a finite space result in discrete solutions that tile space has been known for a long time \cite{Smith1971, Huang2005} and studied in a number of systems from neuroscience \cite{Nikitin2009} to ecology \cite{Johnstone1994}. The only way to obtain a continuous optimal distribution when optimizing information is to consider an unbounded space. However the tiling solution discussed here in terms of gene regulation is slightly different. The genes are already discrete, imposing the discrete structure on the problem. The input concentration range adds an additional layer of potential discreteness (clustering) that is different from the discreteness of the information optimal distributions discussed in other contexts. The discrete number of genes is separated in concentration space, which they tile, building a discrete structure from already discrete units. This type of solution is analogous to optimal tiling of the visual receptive range by retinal ganglion cells  \cite{Borghuis2008,Gauthier2009} and linked to the hypothesis of efficient coding \cite{Barlow1961}. The pattern of repressing neurons that process visual stimuli in the retina ("lateral inhibition") has been proposed as a way to remove redundancy in the encoding of the signal that comes from correlations in the environment (visual stimulus) and the response (receptive field). 

More generally, the appearance of discretized solutions in continuous systems, similar to the tiling patterns described here has been described in many different contexts. Many of these analogies, pointed out to me by Oliviere Rivoire and Thierry Mora, are often linked to bet-hedging strategies \cite{Kelly1956} with applications ranging from phenotypic stability \cite{Sasaki1995}, competition of individuals for resources \cite{Sasaki1997}, ecology models of population regulation \cite{Meszena2006}, neural coding \cite{Nikitin2009} to portfolio risk management \cite{Bouchaud2000}. Of course discrete designs of components of naturally occurring biological systems are very common. For example, in neuroscience both the visual system \cite{masland-2001} and the olfactory system \cite{axel-1991} use discrete receptors to respond to continuous or quasi-continuous inputs, whereas the existence of species is responsible for the interest of ecologists in this topic. The {retina} system has been characterized in great detail experimentally \cite{Shlens2008} and theoretical detailed predictions have been put forward based on the idea of {\it{efficient coding}}, which {hypothesizes} that signal processing has evolved {to optimally encode natural stimulus} while minimizing resources, to the extent that it is becoming possible to test these predictions against existing the neural circuitry \cite{Meister1995, Balasubramanian2002, Schneidman2006, Puchalla2005}.

Lastly, we can ask what is the link between these optimal solutions, which are obtained in a static setting, to real dynamical biological systems. How is the structure of non-overlapping genes and well separated receptors that nevertheless manage to completely cover space achieved in terms of the natural dynamics of the systems? As was explicitly shown in \cite{Mayer2015} a {simple} birth-death dynamics of receptor clones where receptors compete for interactions with antigens results in exactly the same optimal repertoire distributions as discussed above. There, mutual competitive exclusion  is responsible for the characteristic tiling structure.  In genes, the concentration ranges in which genes function {are also mutually exclusive, as evidenced by} their {distinct dissociation} constants. In our information optimization setup, the genes {do not interact}. {Yet} the system achieved this  {solution} by means of long term evolution, which favored this non-overlapping state. In fact, allowing  {two} output genes to interact results in an optimal network where {the gene with a higher dissociation constant represses the expression of the gene with a smaller dissociation constant}. This further restricts the range of activity of the gene {with a lower dissociation constant}, since it gets turned completely off beyond a certain concentration. In this interacting gene example the exclusion is encoded both in the {dissociation} constants and the direct negative {regulation}. In general evolution can use either one of these methods to arrive at non-overlapping concentration ranges where these genes function, hence tiling solutions. Negative feedback or competition is essential for the dynamics to reach these locally clustered, globally tiled solutions also in other systems. Such behavior has been termed competitive exclusion in ecology \cite{Szabo2006} and evoked as a reason for the emergence of species, as well as lateral inhibition in neuroscience \cite{Hartline1956}.

{\bf Acknowledgements} 

Naturally I am very grateful to my collaborators with whom I carried out the work presented in this review: Vijay Balasubramanian, William Bialek, Andreas Mayer, Thierry Mora and Gasper Tkacik. Additionally I thank Thierry Mora and Olivier Rivoire for insightful discussions about the emergence of discrete solutions in various biological systems. I also thank all my past and present collaborators for creating a stimulating environment to consider in detail the functioning of many biological system. The repertoire work was supported by grant ERCStG n. 306312 and I thank the MCCIG PCIG10-GA-2011-303561for support.

\bibliographystyle{pnas}
\bibliography{optimmune,thierry,rev_cras}

\begin{thebibliography}{10}

\bibitem{Hopfield1974}
Hopfield J
\newblock (1974) {Kinetic proofreading: a new mechanism for reducing errors in
  biosynthetic processes requiring high specificity}.
\newblock \emph{Proceedings of the National Academy of Sciences of the United
  States of America} 71:4135--4139.

\bibitem{Ninio1975}
Ninio J
\newblock (1975) {Kinetic amplification of enzyme discrimination}.
\newblock \emph{Biochimie} 57:587--95.

\bibitem{Tostevin2007}
Tostevin F, ten Wolde PR, Howard M
\newblock (2007) {Fundamental limits to position determination by concentration
  gradients.}
\newblock \emph{PLoS computational biology} 3:e78.

\bibitem{saunders2009}
Saunders TE, Howard M
\newblock (2009) {Morphogen profiles can be optimized to buffer against noise}.
\newblock \emph{Physical Review E} 80:041902.

\bibitem{Tkacik2008a}
Tkacik G, Callan CG, Bialek W
\newblock (2008) {Information flow and optimization in transcriptional
  regulation.}
\newblock \emph{Proceedings of the National Academy of Sciences of the United
  States of America} 105:12265--70.

\bibitem{Mehta2009}
Mehta P, Goyal S, Long T, Bassler BL, Wingreen NS
\newblock (2009) {Information processing and signal integration in bacterial
  quorum sensing.}
\newblock \emph{Molecular systems biology} 5:325.

\bibitem{Walczak2010}
Walczak AM, Tka\v{c}ik G, Bialek W
\newblock (2010) {Optimizing information flow in small genetic networks. II.
  Feed-forward interactions}.
\newblock \emph{Physical Review E} 81:041905.

\bibitem{Dubuis2013}
Dubuis JO, Tkacik G, Wieschaus EF, Gregor T, Bialek W
\newblock (2013) {Positional information, in bits.}
\newblock \emph{Proceedings of the National Academy of Sciences of the United
  States of America} 110:16301--8.

\bibitem{Tostevin2009}
Tostevin F, ten Wolde PR
\newblock (2009) {Mutual Information between Input and Output Trajectories of
  Biochemical Networks}.
\newblock \emph{Physical Review Letters} 102:218101.

\bibitem{Tostevin2010}
Tostevin F, ten Wolde PR
\newblock (2010) {Mutual information in time-varying biochemical systems}.
\newblock \emph{Physical Review E} 81:061917.

\bibitem{deRonde2010}
de~Ronde WH, Tostevin F, ten Wolde PR
\newblock (2010) {Effect of feedback on the fidelity of information
  transmission of time-varying signals}.
\newblock \emph{Physical Review E} 82:031914.

\bibitem{Vergassola2007}
Vergassola M, Villermaux E, Shraiman BI
\newblock (2007) {'Infotaxis' as a strategy for searching without gradients.}
\newblock \emph{Nature} 445:406--9.

\bibitem{Siggia2013}
Siggia ED, Vergassola M
\newblock (2013) {Decisions on the fly in cellular sensory systems.}
\newblock \emph{Proceedings of the National Academy of Sciences of the United
  States of America} 110:E3704--12.

\bibitem{Celani2010}
Celani A, Vergassola M
\newblock (2010) {Bacterial strategies for chemotaxis response.}
\newblock \emph{Proceedings of the National Academy of Sciences of the United
  States of America} 107:1391--6.

\bibitem{Francois2004}
Fran\c{c}ois P, Hakim V
\newblock (2004) {Design of genetic networks with specified functions by
  evolution in silico}.
\newblock \emph{\ldots of the National Academy of Sciences \ldots}
  101:580--584.

\bibitem{Francois2007}
Fran\c{c}ois P, Hakim V, Siggia ED
\newblock (2007) {Deriving structure from evolution: metazoan segmentation.}
\newblock \emph{Molecular systems biology} 3:154.

\bibitem{Francois2010}
Fran\c{c}ois P, Siggia ED
\newblock (2010) {Predicting embryonic patterning using mutual entropy fitness
  and in silico evolution.}
\newblock \emph{Development (Cambridge, England)} 137:2385--95.

\bibitem{Gerland2009}
Gerland U, Hwa T
\newblock (2009) {Evolutionary selection between alternative modes of gene
  regulation.}
\newblock \emph{Proceedings of the National Academy of Sciences of the United
  States of America} 106:8841--6.

\bibitem{Savageau1977}
Savageau Ma
\newblock (1977) {Design of molecular control mechanisms and the demand for
  gene expression.}
\newblock \emph{Proceedings of the National Academy of Sciences} 74:5647--5651.

\bibitem{Scott2010}
Scott M, Gunderson CW, Mateescu EM, Zhang Z, Hwa T
\newblock (2010) {Interdependence of cell growth and gene expression: origins
  and consequences.}
\newblock \emph{Science (New York, N.Y.)} 330:1099--102.

\bibitem{Klumpp2008}
Klumpp S, Hwa T
\newblock (2008) {Growth-rate-dependent partitioning of RNA polymerases in
  bacteria.}
\newblock \emph{Proceedings of the National Academy of Sciences of the United
  States of America} 105:20245--50.

\bibitem{Barlow1961}
Barlow H
\newblock (1961) in \emph{Sensory Communication}
\newblock p 217.

\bibitem{Laughlin1981}
Laughlin S
\newblock (1981) {A simple coding procedure enhances a neuron's information
  capacity}.
\newblock \emph{Z. Naturforsch} 36:910--912.

\bibitem{Tkacik2009}
Tka\v{c}ik G, Walczak A, Bialek W
\newblock (2009) {Optimizing information flow in small genetic networks}.
\newblock \emph{Physical Review E} 80:031920.

\bibitem{Tkacik2011}
Tka\v{c}ik G, Walczak AM
\newblock (2011) {Information transmission in genetic regulatory networks: a
  review.}
\newblock \emph{Journal of physics. Condensed matter : an Institute of Physics
  journal} 23:153102.

\bibitem{Tkacik2012}
Tka\v{c}ik G, Walczak AM, Bialek W
\newblock (2012) {Optimizing information flow in small genetic networks. III. A
  self-interacting gene}.
\newblock \emph{Physical Review E} 85:041903.

\bibitem{Mayer2015}
Mayer A, Balasubramanian V, Mora T, Walczak AM
\newblock (2015) {How a well-adapted immune system is organized}.
\newblock \emph{Proceedings of the National Academy of Sciences USA} pp 1--15.

\bibitem{Lawrence1992}
Lawrence P
\newblock (1992) \emph{{The Making of a Fly: The Genetics of Animal Design.}}
\newblock (Blackwell Scientific Publications, Oxford).

\bibitem{Crauk2005}
Crauk O, Dostatni N
\newblock (2005) {Bicoid determines sharp and precise target gene expression in
  the Drosophila embryo.}
\newblock \emph{Current biology : CB} 15:1888--98.

\bibitem{Delbruck1940}
Delbr\"{u}ck M
\newblock (1940) {Statistical fluctuation in autocatalytic reactions.}
\newblock \emph{J Chem Phys} 8:120--124.

\bibitem{Elowitz2002}
Elowitz MB, Levine AJ, Siggia ED, Swain PS
\newblock (2002) {Stochastic gene expression in a single cell.}
\newblock \emph{Science (New York, N.Y.)} 297:1183--6.

\bibitem{Ozbudak2002}
Ozbudak EM, Thattai M, Kurtser I, Grossman AD, van Oudenaarden A
\newblock (2002) {Regulation of noise in the expression of a single gene.}
\newblock \emph{Nature genetics} 31:69--73.

\bibitem{Raser2004}
Raser J, O'Shea E
\newblock (2004) {Control of stochasticity in eukaryotic gene expression}.
\newblock \emph{Science} 304:1811.

\bibitem{Kaern2005}
Kaern M, Elston TC, Blake WJ, Collins JJ
\newblock (2005) {Stochasticity in gene expression: from theories to
  phenotypes.}
\newblock \emph{Nature reviews. Genetics} 6:451--64.

\bibitem{Walczak2005a}
Walczak AM, Onuchic JN, Wolynes PG
\newblock (2005) {Absolute rate theories of epigenetic stability.}
\newblock \emph{Proceedings of the National Academy of Sciences of the United
  States of America} 102:18926--31.

\bibitem{Hornos2005}
Hornos JEM, {et~al.}
\newblock (2005) {Self-regulating gene: An exact solution}.
\newblock \emph{Physical Review E} 72:051907.

\bibitem{Walczak2012}
Walczak A, Mugler A, Wiggins C
\newblock (2012) in \emph{Computational Modeling of Signaling Networks}
\newblock pp 273--322.

\bibitem{Golding2005}
Golding I, Paulsson J, Zawilski SM, Cox EC
\newblock (2005) {Real-time kinetics of gene activity in individual bacteria.}
\newblock \emph{Cell} 123:1025--36.

\bibitem{Cai2006}
Cai L, Friedman N, Xie XS
\newblock (2006) {Stochastic protein expression in individual cells at the
  single molecule level.}
\newblock \emph{Nature} 440:358--62.

\bibitem{Elf2007}
Elf J, Li GW, Xie X
\newblock (2007) {Probing Transcription Factor Dynamics at the Single Molecule
  Level in a Living Cell}.
\newblock \emph{Science} 316:1191--1195.

\bibitem{Jaeger2011}
Jaeger J
\newblock (2011) {The gap gene network.}
\newblock \emph{Cellular and molecular life sciences : CMLS} 68:243--74.

\bibitem{Shannon1948}
Shannon C
\newblock (1948) {A mathematical theory of communication}.
\newblock \emph{Bell Sys Tech J} 27:623.

\bibitem{Cover1991}
Cover T, Thomas J
\newblock (1991) \emph{{Elements of Information Theory}}
\newblock (John Wiley, New York, New York, USA).

\bibitem{Berg1977}
Berg HC, Purcell EM
\newblock (1977) {Physics of chemoreception.}
\newblock \emph{Biophysical journal} 20:193--219.

\bibitem{Janeway}
Murphy K, Travers P, Walport M
\newblock (2001) \emph{Janeway's Immunobiology}
\newblock (Garland Science) Vol.{}~2, 7th edition.

\bibitem{oster-1979}
Perelson AS, Oster GF
\newblock (1979) Theoretical studies of clonal selection: minimal antibody
  repertoire size and reliability of self-non-self discrimination.
\newblock \emph{Journal of theoretical biology} 81:645--670.

\bibitem{Murugan2012}
Murugan A, Mora T, Walczak AM, Callan CG
\newblock (2012) {Statistical inference of the generation probability of T-cell
  receptors from sequence repertoires.}
\newblock \emph{Proceedings of the National Academy of Sciences of the United
  States of America} 109:16161--6.

\bibitem{Elhanati2014}
Elhanati Y, Murugan A, Callan CG, Mora T, Walczak AM
\newblock (2014) {Quantifying selection in immune receptor repertoires.}
\newblock \emph{Proceedings of the National Academy of Sciences of the United
  States of America} 111:9875--80.

\bibitem{Venturi2006}
Venturi V, {et~al.}
\newblock (2006) {Sharing of T cell receptors in antigen-specific responses is
  driven by convergent recombination.}
\newblock \emph{Proceedings of the National Academy of Sciences of the United
  States of America} 103:18691--6.

\bibitem{Venturi2008}
Venturi V, Price Da, Douek DC, Davenport MP
\newblock (2008) {The molecular basis for public T-cell responses?}
\newblock \emph{Nature reviews. Immunology} 8:231--8.

\bibitem{Mora2010}
Mora T, Walczak AM, Bialek W, Callan CG
\newblock (2010) {Maximum entropy models for antibody diversity.}
\newblock \emph{Proceedings of the National Academy of Sciences of the United
  States of America} 107:5405--10.

\bibitem{Weinstein2009}
Weinstein Ja, Jiang N, White Ra, Fisher DS, Quake SR
\newblock (2009) {High-throughput sequencing of the zebrafish antibody
  repertoire.}
\newblock \emph{Science (New York, N.Y.)} 324:807--10.

\bibitem{Fairhall2001}
Fairhall A, Lewen G, Bialek W, van Steveninck R
\newblock (2001) {Efficiency and ambiguity in an adaptive neural code}.
\newblock \emph{Nature} 412:787.

\bibitem{Andrews2007}
Andrews BW, Iglesias Pa
\newblock (2007) {An information-theoretic characterization of the optimal
  gradient sensing response of cells.}
\newblock \emph{PLoS computational biology} 3:e153.

\bibitem{Smith1971}
Smith JG
\newblock (1971) {The information capacity of amplitude- and
  variance-constrained sclar gaussian channels}.
\newblock \emph{Information and Control} 18:203--219.

\bibitem{Huang2005}
Huang J, Meyn S
\newblock (2005) {Characterization and Computation of Optimal Distributions for
  Channel Coding}.
\newblock \emph{IEEE Transactions on Information Theory} 51:2336--2351.

\bibitem{Nikitin2009}
Nikitin A, Stocks N, Morse R, McDonnell M
\newblock (2009) {Neural Population Coding Is Optimized by Discrete Tuning
  Curves}.
\newblock \emph{Physical Review Letters} 103:138101.

\bibitem{Johnstone1994}
Johnstone R
\newblock (1994) {Honest signalling, perceptual error and the evolution of
  "all-or-nothing" displays}.
\newblock \emph{Proceedings of the Royal Society B: Biological Sciences}
  256:169--175.

\bibitem{Borghuis2008}
Borghuis BG, Ratliff CP, Smith RG, Sterling P, Balasubramanian V
\newblock (2008) {Design of a neuronal array.}
\newblock \emph{The Journal of neuroscience : the official journal of the
  Society for Neuroscience} 28:3178--89.

\bibitem{Gauthier2009}
Gauthier JL, {et~al.}
\newblock (2009) {Receptive fields in primate retina are coordinated to sample
  visual space more uniformly.}
\newblock \emph{PLoS biology} 7:e1000063.

\bibitem{Kelly1956}
Kelly, J. J
\newblock (1956) {A new interpretation of information rate}.
\newblock \emph{IRE Transactions on Information Theory} 2:917--926.

\bibitem{Sasaki1995}
Sasaki A, Ellner S
\newblock (1995) {The evolutionary stable phenotype distribution in a random
  environment}.
\newblock \emph{Evolution} 49:337--350.

\bibitem{Sasaki1997}
Sasaki A
\newblock (1997) {Clumped Distribution by Neighborhood Competition}.
\newblock \emph{J. theor. Biol.} 186:415--430.

\bibitem{Meszena2006}
Mesz\'{e}na G, Gyllenberg M, P\'{a}sztor L, Metz JaJ
\newblock (2006) {Competitive exclusion and limiting similarity: a unified
  theory.}
\newblock \emph{Theoretical population biology} 69:68--87.

\bibitem{Bouchaud2000}
Bouchaud J, Potters M
\newblock (2000) \emph{{Theory of Financial Risks}}
\newblock (Cambridge University Press, Cambridge, UK).

\bibitem{masland-2001}
Masland RH
\newblock (2001) The fundamental plan of the retina.
\newblock \emph{Nature Neuroscience} 4:877--886.

\bibitem{axel-1991}
Buck L, Axel R
\newblock (1991) A novel multigene family may encode odorant receptors: A
  molecular basis for odor recognition.
\newblock \emph{Cell} 65:175--187.

\bibitem{Shlens2008}
Shlens J, Rieke F, Chichilnisky E
\newblock (2008) {Synchronized firing in the retina.}
\newblock \emph{Current opinion in neurobiology} 18:396--402.

\bibitem{Meister1995}
Meister M, Lagnado L, Baylor DA
\newblock (1995) {Concerted Signaling by Retinal Ganglion Cells}.
\newblock \emph{Science} 8:1207--1210.

\bibitem{Balasubramanian2002}
Balasubramanian V, Berry M
\newblock (2002) {A test of metabolically efficient coding in the retina}.
\newblock \emph{Network: Computation in Neural Systems} 13:531--552.

\bibitem{Schneidman2006}
Schneidman E, Berry MJ, Segev R, Bialek W
\newblock (2006) {Weak pairwise correlations imply strongly correlated network
  states in a neural population.}
\newblock \emph{Nature} 440:1007--12.

\bibitem{Puchalla2005}
Puchalla JL, Schneidman E, Harris Ra, Berry MJ
\newblock (2005) {Redundancy in the population code of the retina.}
\newblock \emph{Neuron} 46:493--504.

\bibitem{Szabo2006}
Szab\'{o} P, Mesz\'{e}na G
\newblock (2006) {Limiting similarity revisited}.
\newblock \emph{Oikos} 3:612--619.

\bibitem{Hartline1956}
Hartline HK, Wagner HG, Ratliff F
\newblock (1956) {Inhibition in the eye of limulus}.
\newblock \emph{The Journal of General Physiology} 39:651--673.

\end{thebibliography}

\end{document}